# Tailoring Zr-Based Metal–Organic Frameworks by Decorating UiO-66 with Amino Group for Adsorption Removal of Perfluorooctanoic Acid from Water

Sicheng Ding and Xiaolan Zeng

## Abstract

As a persistent pollutant, PFOA places a serious threat to natural ecosystems and human health for its global distribution and residual. Currently, adsorption technology is a promising approach for PFOA removal. MOFs become a concerned adsorbent attributing to its special designability and easy modular synthesis. Aiming at improving hydrophobic interactions to absorb PFOA, this study was designed to introduce amino group to decorate Zr-based MOF UiO-66. And subsequently introduced GO to modify UiO-66-NH$_2$ to explore the influence of hydrophobicity on PFOA adsorption removal. Based on the performance of the characterization of synthesized products, the adsorption removal effect of PFOA and the corresponding adsorption mechanism were discussed. The results indicate the introduction of the amino group could enlarge the pore size, but also display strong hydrophobic effects, significantly enhance the adsorption capacity(maximum capacity was 1653mg/g). This work not only provides the method of introducing amino group to Zr-based MOFs to strengthen the hydrophobic properties in purpose of removing PFOA efficiently, but also proposes the feasible the reference for tailoring MOFs to remove endurable organic pollutants in water.

**Graphical abstract:**

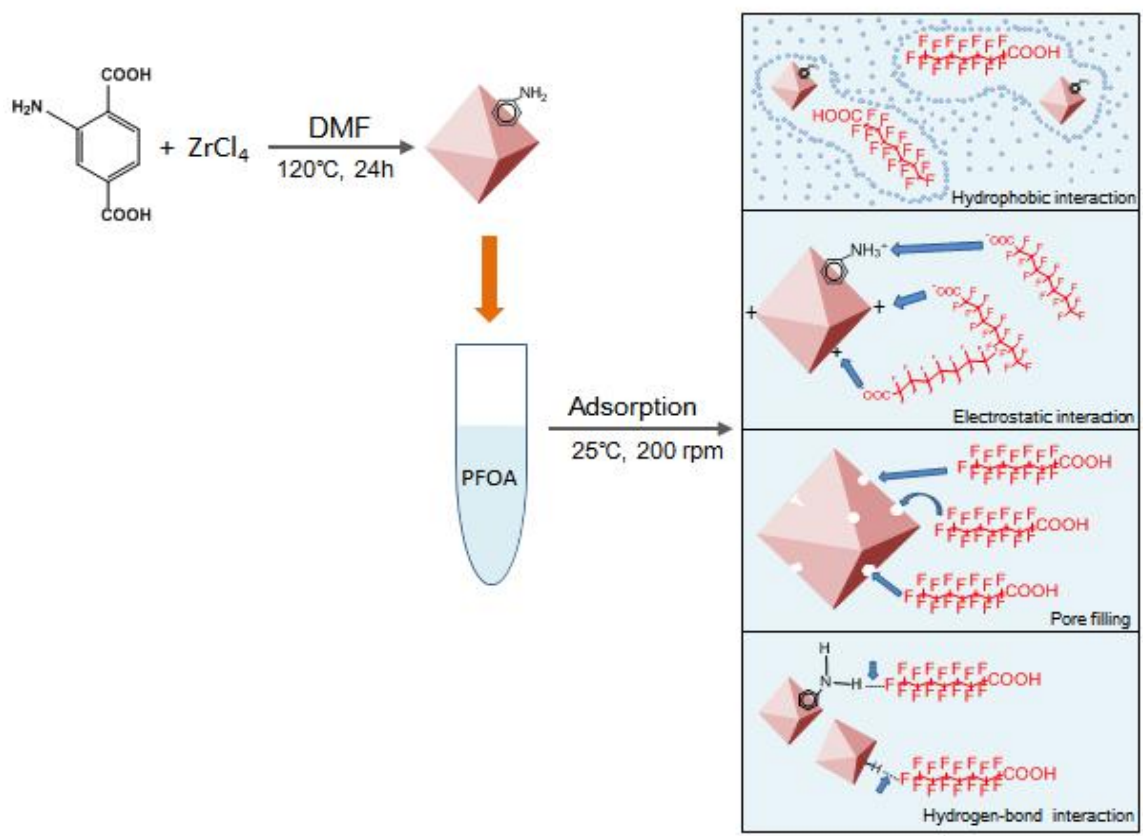

**Table of Contents**

# 1 Introduction

## 1.1 Properties, applications, and hazards of perfluorooctanoic acid

As the most concerned typical per- and polyfluorinated chemicals (PFCs)[1], pentadecafluorooctanoic acid (PFOA) is artificially synthesized with an eight-carbon chain, 15 carbon-fluorine bonds and a carboxyl group as shown in Fig.1[2]. The carbon-fluorine bonds in PFOA are strongly polarized and have high bond energy (460kJ·mol$^{-1}$)[3] due to the strong electronegativity(-4.0) of the fluorine atom. That results in the high physiochemical stability of PFOA to be resistant to strong acid and oxidant. Meanwhile, relatively low polarizability of carbon-fluorine bond and molecular cohesion in PFOA contributes to good hydrophobicity and lipophobicity[4]. In addition, PFOA also has high surface activity and strong perseverance against light, heat, microbial degradation and biological metabolism. Featured by all these properties, PFOA has been manufactured world-wide since the 1950s and prepared for extensive applications in daily products and manufacturing, including textile, paper, leather, foamite, medicine, cosmetics, and food packaging, etc.[5].

Because of the wide application and diverse transportation routes, bioaccumulation, and degradation-resistant, as well as being the metabolic end product of some perfluorinated compounds and precursor substances such as fluorotelomer alcohols (FTOHs) in the environment and organisms, PFOA is globally distributed and residual[6]. At present, PFOA could be found in almost all environmental media such as water, soil, atmosphere, dust, sediments, animals, animals and human bodies [7], and it has become the perfluorinated compound with the highest concentration and detection rate in the environment. As a nonvolatile and strong organic acid, PFOA mainly exists in water with solubility of 3.4g/L at 20°C and almost complete ionization (pKa = 2.5)[8].

Massive researches reveal that PFOA places a serious threat to natural ecosystems and human health as a persistent pollutant that possesses multiple toxicities such as heredity, immunity, reproduction, development, and nerves, as well as causing severe dysfunctions in tissues and organs in organisms. In the wake of the increasing usage of PFOA and its precursors, their increasingly serious toxicity causes environment contamination worldwide. The United States Environmental Protection Agency (USEPA) cataloged the PFOA as a carcinogen [9]. It has been restricted or banned to use not only in the United States, but also in Canada and Germany[10], where all emissions of PFOA would be ceased with the agreement of manufacturers(FluoroCouncil members). Compared with its precursors, PFOA has aroused extensive interests from researchers with a longer half-life and stronger toxicity[9].

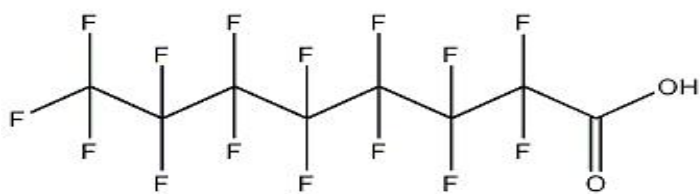

Fig. 1  Molecular structure of PFOA

Table 1  Physical and chemical properties of PFOA

| PFCs | Molecular formula | solubility (20 ℃) | molecular weight (Da) | fusion point (℃) | pKa |
| --- | --- | --- | --- | --- | --- |
| PFOA | $C_8HF_{15}O_2$ | 3400 mg·$L^{-1}$ | 414.07 | 55~60 | 2.5 |

Note: pKa is the dissociation constant.

## 1.2 Removal of PFOA in water environment

Recently, many physical and chemical technologies have been developed to eliminate PFOA from environment such as coagulation, electromembrane process, photochemical oxidation, photocatalysis, microwave degradation, electrochemical degradation and adsorption technology[9].Since low cost, easy operation, less energy requirement and loose reaction conditions, adsorption technology has attracted great attention[11].To date, a variety of porous materials have been developed to adsorb PFOA from wastewater, such as mesoporous metal oxides (silica and alumina), microporous minerals (zeolite) and porous carbons (activated carbon, carbon nanotubes and carbon fiber)[9]. Generally, the research results suggested electrostatic interactions, acid-base interactions, pore filling adsorption, hydrogen bonding, hydrophobic interaction and π-π interaction as the main reaction mechanisms [11]. Current studies on PFOA adsorption by biochar have shown that the van der Waals and π-π bonds would negligibly affect the adsorption process considering the low polarizability and absence of benzene structure in PFOA [12]. Thus promoting the electrostatic and hydrophobic interaction between adsorbent and PFOA would be a possibility to improve the adsorption. In decision for potential absorbing materials, factors above that would affect the adsorption mechanism should be the priority.

## 1.3 Metal-organic framework material and application

Metal-organic frameworks (MOFs) ,fabricated by metal cluster and organic ligands, are of hybridized organic-inorganic compounds with micro/meso- pores structure [13]. They can be divided into IRMOFs, ZIFs, MILs, UIO, HKUST-1, POST-1, NOTT-300, etc. based on different component units and synthesis methods. MOFs show some superior characteristics including high specific surface area, abundant unsaturated metallic sites, and high mechanical stability, as well as easy modular synthesis, designable pore structure, and tunable physicochemical properties targeting different adsorbents via variation of metal ions and organic ligands or decoration on synthesized products[14]. In recent decades, MOFs have shown great separation and adsorption potential[11], and extensively applied to storage/separation, catalysis, sensing, bioimaging and other aspects[13]. At present, MOFs are widely used in environmental engineering[10], and in particular, its composite materials have been absorbents for various contaminants in water [13]. Most researches have been conducted to increase selective adsorption efficiency of MOFs by post-synthetic modification on the basis of adsorption mechanisms. Post-synthetic modification means supplementing new functional groups, introducing new chemical activated surface sites[15] or functional adjustments of synthesized MOFs[16].

**1.4 Zr-based metal organic framework material and application**

The Zr-based MOF UiO-66 is a porous material with 11 Å Zr octahedra cage and an 8 Å highly packed face-centered cubic (FCC) crystal structure. These different cage structures are connected through each other to form a triangle-shaped structure whose width is close to 6Å so that only small molecules could pass through. Due to very strong Zr–O bond and dense clusters, UiO-66 and its derivatives possess unprecedented high mechanical, chemical and thermal stability. They can withstand harsh aqueous conditions such as strong basic and acidic solutions, organic solvents and aqueous solution, and can maintain crystallinity after aging in water for 12 months. Compared with many MOFs only lasting for a shorter period of time, they are few known water-stable MOFs and easy to be directly modified by ligand substitution[17]. In this case, this family of material has attracted widespread attention in gas adsorption, drug sustained-release, heterogeneous catalytic, fuel cell and other aspects [16].

It has been reported that asymmetric coordination of the amino group facilitated to engineer larger pores size, surface size and higher zeta potential on UiO-66 surface having –$NH_2$ groups, which are awarded high kinetics towards anion[13]. Under hydrogen bonding, UiO-66 decorated with the amino group has significant absorption capacity of sulfachloropyridazine(SCP)[13]. UiO-66-$NH_2$ demonstrated the highest potential to adsorb phenol red (PR) through acid-base interaction, hydrogen bonding and π-π interaction between dye and sorbent molecules [11]. In addition, graphene oxide-containing compounds indicated application in actual wastewater containing organic or inorganic pollutants treatment [18]. Furthermore, GO can easily bind with metallic clusters of MOFs owing to its electronegativity. And the large surface area, aromatic sp$^2$ domain with epoxy and carboxylic groups in GO can play an important role in the bonding interaction and growth of MOFs[19]. Recent research indicates that the removal efficiency of dyes and antibiotics in wastewater with the original UiO-66-(OH)$_2$ can be improved effectively by GO modification, which is attributed to the addition of more active adsorption sites, improving the stability and space utilization of the material[18].

### 1.5 The target and significance of the research

At present, few researches on the application of MOFs to remove PFOA from water have been reported. Liu et al.[7] utilized Cr-MIL-101 to obtained high PFOA removal due to its high surface area and large pore volume. Chen et. al.[7] developed zeolitic imidazolate frameworks to remove PFOA. Yang et al.[9] investigated the interfacial interaction mechanism of PFOA adsorption for Fe-based MOFs including Fe-BTC, MIL-100-Fe and MIL-101-Fe using experiments and computational calculation at molecular level even to electronic level; Sini et al.[20] studied the adsorption of PFOA for Zr-based UiO-66 and equivalent perfluorinated UiO-66-F$_4$. These researches mainly focus on exploring the effects and mechanisms for the typical MOFs adsorption of PFOA, rather than increasing the adsorption capacity through modifying with new functional groups on MOFs aiming to the basic properties of PFOA.

Based on previous studies of adsorption removal techniques of PFOA in water environment combined with the contemporary progress of the UiO-66 in adsorption field, targeting good hydrophobicity of PFOA and its anionic existence (C$_7$F$_{15}$COO$^-$) under pH=2~10[12], this study is designed to introduce amino group to decorate UiO-66 aiming at improving electrostatic interaction and hydrophobic interactions. To explore the influence of hydrophobicity on PFOA adsorption removal, GO, a rich oxygen-containing groups was further introduced to modify UiO-66-NH$_2$. Based on the performance characterization of UiO-66 and its derivatives, the adsorption effect of PFOA and the corresponding adsorption mechanism were discussed to identify the role of amino group on UiO-66 for the impact of adsorption performance. The results proposed the feasible the reference for tailoring MOFs to remove endurable organic pollutants in water.

## 2 Experimental

### 2.1 Materials and equipment

Table 2 and Table 3 list the major chemicals, reagents and equipment in this research. Part of the chemicals and equipment are displayed in Fig. 2. All chemicals were received without further purification.

Table 2  Main materials in the experiment

| chemical name | molecular formula | purity | manufacturer |
| --- | --- | --- | --- |
| Perfluorooctanoic acid | $C_8HF_{15}O_2$ | Chromatographically | Shanghai Aladdin Biochemical Technology Co., Ltd. |
| ZirconiumTetrachloride | $ZrCl_4$ | Analytical | Shanghai Aladdin Biochemical Technology Co., Ltd. |
| Terephthalic acid | $C_8H_6O_4$ | Analytical | Shanghai Aladdin Biochemical Technology Co., Ltd. |
| 2-aminoterephthalic acid | $C_8H_7NO_4$ | Analytical | Shanghai Aladdin Biochemical Technology Co., Ltd. |
| Graphene Oxide | $C_{140}H_{42}O_{20}$ | Analytical | ShangHai YuanYe Biotechnology Co., Ltd |
| Dimethylformamide | $C_3H_7NO$ | Analytical | Chengdu Kelong Chemical Co., Ltd. |
| Acetic Acid | $CH_3COOH$ | Analytical | Chongqing Chuandong Chemical (Group) Co., Ltd. |
| Methanol | $CH_3OH$ | Analytical | Shanghai Aladdin Biochemical Technology Co., Ltd. |
| Ethanol | $CH_3CH_2OH$ | Analytical | Chongqing Chuandong Chemical (Group) Co., Ltd. |

Table 3  Major equipment in the experiment

| equipment name | manufacturer | model |
| --- | --- | --- |
| Electronic Balance | Techcomp Precision Balances (Shanghai) Co., Ltd | FA2004B |
| Ultrapure Water System | Sichuan Water Purifier Treatment Equipment Co., Ltd | WP-UPT-10 |
| Ultrasonic Cleaner | NingBo Scientz Biotechnology Co.,Ltd | SB25-12D |
| Stainless Reaction Vessel with a Tetrafluoroethylene Liner | Beijing Getimes Technology Co., Ltd | four 20mL vessels, one 100mL and 250mL vessel |
| Blast Drying Oven | Shanghai Boxun Industry & Commerce Co.,Ltd | GZX-9030MBE |
| High-speed Centrifuge | Sichuan Shuke Yiqi Co.,Ltd | TG-16 |
| Vacuum Oven | Shanghai Langgan Shiyan Shebei Co.,Ltd | DZF-6050 |
| Liquid Chromatography–Mass Spectrometer | Thermo Fisher Scientific Inc. | Q Exactive Plus |
| Environmental Scanning Electron Microscope | Thermo Fisher Scientific Inc. | Quattro S |
| Powder X-ray Diffractometer | Spectris plc. | PANalytical X'Pert |

| | | Powder |
| --- | --- | --- |
| Fourier-Transform Infrared Spectrometer | Thermo Fisher Scientific Inc. | Nicolet iS50 |
| Automated Fully MultiStation Specific Surface Area and Porosity Analyzer | Anton Paar QuantaTec Inc. | Quadrasorb 2MP |
| Water Contact Angle Goniometer | KRÜSS Scientific Instrument Inc. | DSA JY-82B |
| Zeta Potential Analyzer | Brookhaven Instrument Corp. | NanoBrook Omni |

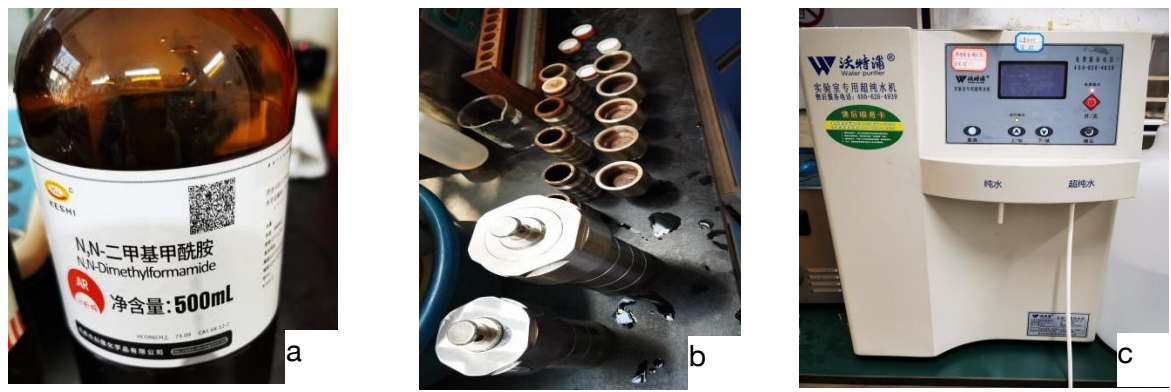

**Fig. 2  Part of the chemicals and equipment in the experiment (a. Dimethylformamide; b. Stainless Reaction Vessel with Tetrafluoroethylene Liners; c. Ultrapure Water System)**

### 2.2 Preparation of Zr-based metal organic framework

Based on previous reported procedures[21,22], UiO-66 and its modified materials were prepared through a simple hydrothermal process with some improvements. The synthesis methods of UiO-66, UiO-66-NH$_2$ and UiO-66-NH$_2$/GO employed in this paper were described as follows. Parts of the synthesis process are shown in Fig. 3 and synthesized products are shown in Fig. 4.

#### 2.2.1 Synthesis of UiO-66

2.16mmol of ZrCl$_4$ and 2.16mmol of terephthalic acid were dissolved in 480mL of DMF. Acetic acid (29mL) was mixed with above mixture. After 10 min sonication, the solution was transferred into the reaction vessels. Then the reactor was placed in the blast drying oven and heated at 120˚C for 24 hours. The prepared product was successively centrifuged under the speed of 8000 rounds per minute(rpm) for 10min. Over the sufficient separation, washed the precipitate with DMF and methanol in a volumetric ratio of 4:1 and followed in a vacuum oven drying at 120˚C. Finally, obtained UiO-66 MOFs.

#### 2.2.2 Synthesis of UiO-66-NH$_2$

Following 2.2.1, ZrCl$_4$ and 2-aminoterephthalic acid dissolved in DMF was added by acetic acid for 10 min sonication. Then, the dark yellow solution was transferred into the reaction vessels with tetrafluoroethylene liners, and placed vessels in the blast drying oven and heated at 120°C for 24 hours, then cooled at room temperature. The prepared product was centrifuged for 10min. Over the sufficient separation, washed the precipitate with DMF and methanol in a volumetric ratio of 4:1 and followed in a vacuum oven drying at 120°C. Finally, obtained UiO66-NH$_2$ nanocomposites.

### 2.2.3 Synthesis of UiO-66-NH$_2$/GO

At first, dissolved UiO-66-NH$_2$ （0.050mmol） yellow powder and 3mL graphene oxide in DMF (480mL). Then, 29mL acetic acid was added into the above mixture after 10 min sonication, added. Then, transferred the black solution into the reaction vessels with tetrafluoroethylene liners, and placed vessels in the blast drying oven and heated at 120°C for 24 hours, then cooled at room temperature. Successively, centrifuged the prepared product under the speed of 8000rpm. The precipitate was washed with anhydrous ethanol and dried at 120°C in a vacuum oven to obtain UiO-66-NH$_2$/GO nanocomposites.

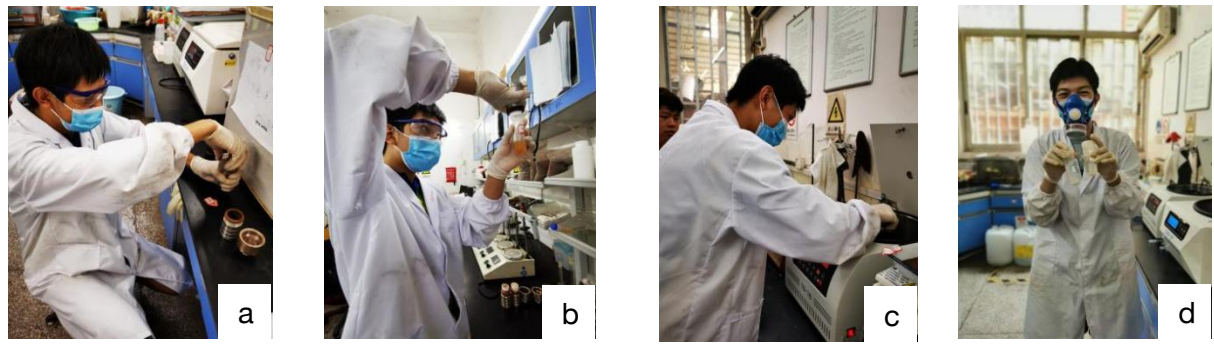

Fig. 3 Part of the synthesis process for Zr-based MOFs (a. Opening the cooled reaction vessel; b. Transferring the mixed solution; c. Centrifuging d. Obtaining the Products)

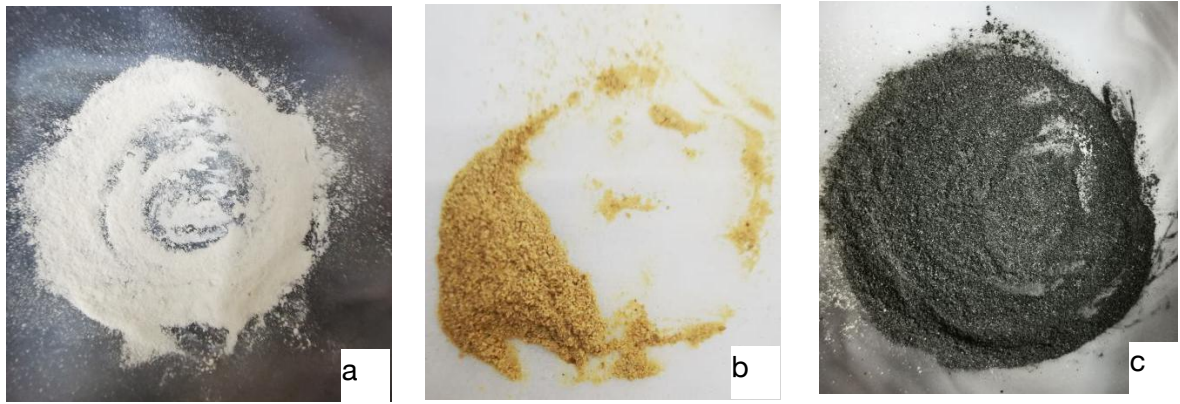

Fig. 4 Synthesized products (a. UiO-66; b.UiO-66-NH$_2$; c. UiO-66-NH$_2$/GO)

## 2.3 Characterization

The characterization methods for each synthesized product are described below, and the determinations of some performance indicators were shown in Fig. 5.

### 2.3.1 X-ray Diffraction (PXRD)

X-ray diffraction patterns of the prepared materials were determined by a Powder diffractometer with Cu Kα(λ=0.154 nm) radiation with a scanning range of 2θ=5°~90°.

### 2.3.2 Environmental Scanning Electron Microscopy (SEM)

The environmental scanning electron microscopy on the Quattro S microscope was used to probe the surface morphology of samples after the sample was sprayed with gold for 10s.

### 2.3.3 Fourier-Transform Infrared Spectrometry (FT-IR)

Fourier-Transform Infrared Spectroscopy (FT-IR) was used to detect the surface chemical groups of samples by the Nicolet iS50 spectrometer using KBr support method. The spectra were scanned from 350-7800 cm$^{-1}$ with a speed of 0.158-6.28 cm•s$^{-1}$.

### 2.3.4 Nitrogen Adsorption-Desorption Isotherms

The Brunauer-Emmett-Teller (BET) surface areas and pore diameter of the samples were calculated by nitrogen adsorption and desorption at -191.15°C(77K) by a fully automated multi-station specific surface area and porosity analyzer( Quadrasorb 2MP). The samples were analyzed at 200°C under high vacuum.

### 2.3.5 Contact Angle

The hydrophilic and hydrophobic properties of synthesized samples were measured by the water contact angle analyzer DSA JY-82B.

### 2.3.6 Zeta Potential

The zeta potential of samples was measured by a NanoBrook Omni Zeta potential analyzer.

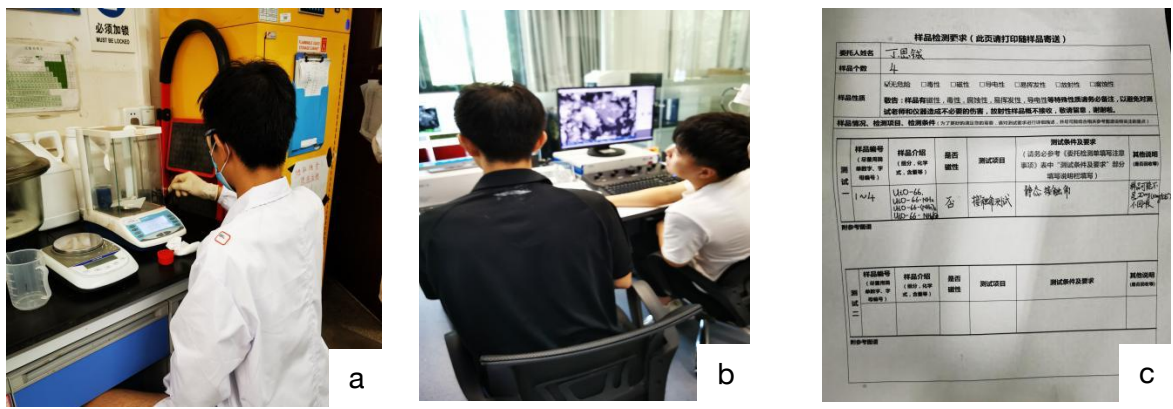

**Fig. 5 Determinations of some performance indicators**
(a. Weighing samples; b. SEM measurements; c. Contact Angle sample test request sheet)

### 2.4 Absorption experiment

The experimental procedure for adsorption kinetics and isotherms for each synthesized product is as the following. Part of the experimental process was displayed in Fig. 6.

### 2.4.1 Adsorption kinetics

The predetermined 0.01g of UiO-66, UiO-66-NH$_2$ and UiO-66-NH$_2$/GO were added into centrifuge tubes with 50 mL of 0.1g/L PFOA solution, respectively. The pH of the solution was adjusted to 3±0.1 with 4M NaOH and 1M HCl solution. The mixture was placed in the incubator shaker for adsorption kinetics experiments with a fixed shaking rate of 200 rpm and temperature of 25°C. At 0, 10, 20, 30, 90, 180, 720, 960, 1440 minutes, diluted the mixture(1.0mL) to 1/100 of its concentration respectively. And then took 1.5mL of each diluted solution to filtrate with a 0.22μm membrane filter (polyethersulfone membrane). Finally, collected the filtered solution with plastic bottle and stored them under 4°C for ultra-high performance liquid chromatography/mass spectrum (UPLC/MS) analysis.

### 2.4.2 Adsorption isotherms

The exact amount of 0.01g of UiO-66, UiO-66-NH$_2$ and UiO-66-NH$_2$/GO into 7 centrifuge tubes with 50 mL of 0.005g/L, 0.015g/L, 0.03g/L, 0.05g/L, 0.1g/L, 0.2g/L and 0.3g/L PFOA solution. The pH of the solution was adjusted to 3±0.1. Then placed the above mixture into the incubator shaker with 200 rpm at 25°C for isothermal sorption experiments. When the adsorption process reached the equilibrium, each mixture(1.0mL) was diluted to 1/100 of its concentration, and then 1.5mL diluted solution was taken to filtrate with a 0.22μm membrane filter (polyethersulfone membrane). The filtered solution was collected for further UPLC/MS analysis.

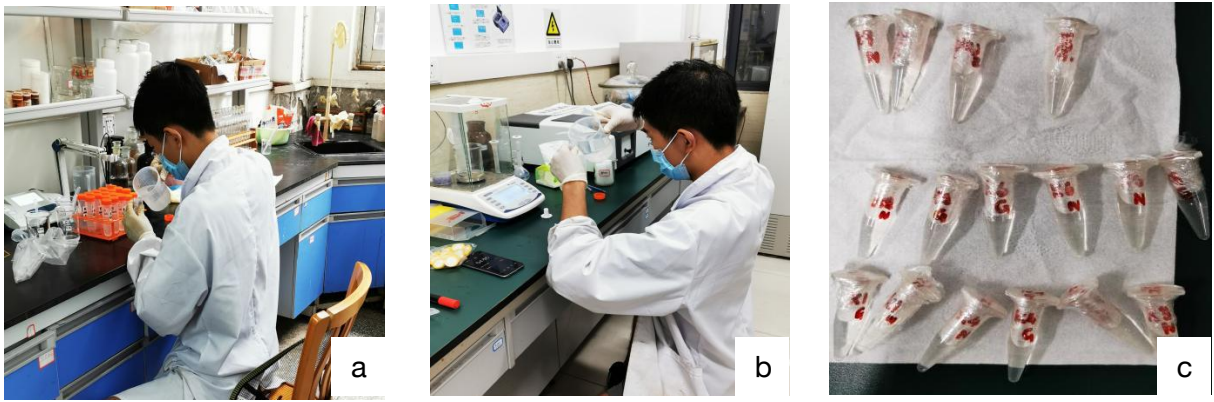

**Fig. 6  Part of experiment procedure for adsorption process (a. Sample preparation for adsorption kinetics; b. Dilution and filtration for samples; c. Water samples for PFOA test)**

## 3 Results and discussion

### 3.1 Characterization analysis

#### 3.1.1 PXRD analysis

The PXRD results of UiO-66, UiO-66-NH$_2$ and UiO-66-NH$_2$/GO are shown in Fig.7. Densest characteristic peaks of UiO-66 occurred at 2θ=7.3°, 8.5° and 25.8° are related to (111), (200) and (224) crystal planes[18]. Meanwhile, the typical peak positions of UiO-66-NH$_2$ are the same as UiO-66, which is in accordance with the previous publications [23]. The results suggest that UiO-66-NH$_2$ material was successfully prepared, and the inherent crystal structure of the original UiO-66 was not influenced by the amino group. It is noticed that the position of peaks of UiO-66-NH$_2$/GO are similar to UiO-66-NH$_2$. It indicates that the introduction of GO did not change the intrinsic crystal structure of UiO-66-NH$_2$ matrix except some peaks were weakened. It should be noted that the characteristic peak (2θ = 9.95°, 11° and 42°) of GO did not appear in UiO-66-NH$_2$/GO composites, suggesting that low contents of GO were highly dispersed by intercalating UiO-66-NH$_2$ particles[18-19].

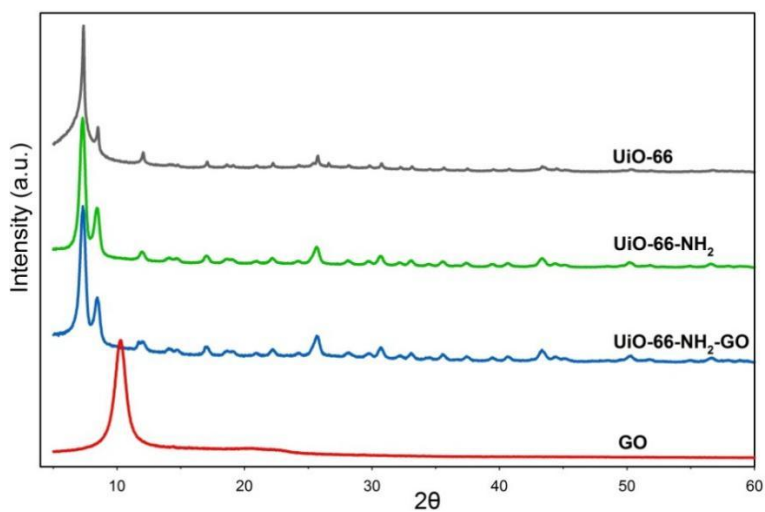

Fig.7　PXRD patterns of materials

### 3.1.2 SEM analysis

The morphology of four MOFs was characterized by SEM (Fig. 8 (a) ~ (d)). It can be seen that GO exhibits irregular sized layer structure and UiO-66 demonstrates agglomerated cubic shapes with similar dimensions. For UiO-66-NH$_2$, even modified by amino group, it also has homogeneous intergrowth cubic crystals, which is the same morphology of UiO-66. Nevertheless, UiO-66-NH$_2$ scatters randomly and accumulates, having a different and smaller size[24]. And the structure of UIO-66-NH$_2$/GO clearly differentiates from the typical agglomerated cubic shape of UiO-66 due to GO decoration, which resulting in larger size and more uneven surface of the particles [19]. Therefore, presence of GO and amino groups increased the irregularities of the material, which created a different microstate and size of UiO-66-NH$_2$ and UiO-66-NH$_2$/GO[18].

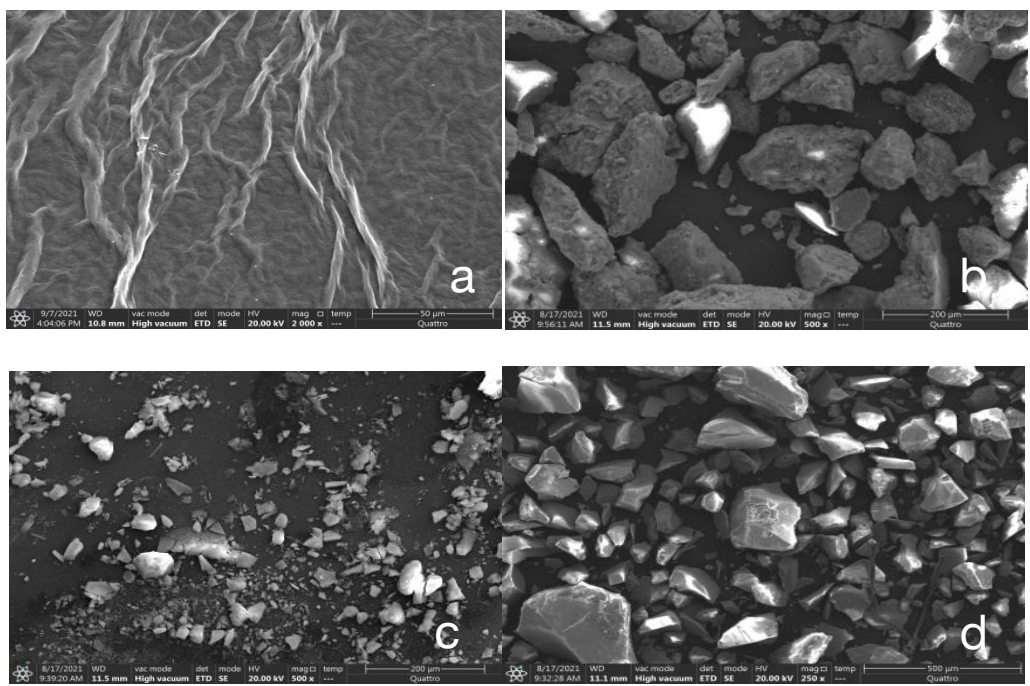

**Fig. 8** SEM images of synthesized products (a. GO; b. UiO-66; c. UiO-66-NH$_2$; d. UiO-66-NH$_2$/GO)

### 3.1.3 FT-IR analysis

In this research, FT-IR spectroscopy was employed to confirm the successful successive introduction of amino and GO functional groups into UiO-66[25]. The FTIR spectra of four materials were demonstrated in Fig. 9. Two characteristic peaks at 700~900 cm$^{-1}$ and at about 660 cm$^{-1}$, which can be observed in all three Zr-Based MOFs referring to the Zr–O stretching vibration, respectively [11,19]. In addition, two strong peaks observed at 1400 and 1590 cm$^{-1}$ and one small peak at about 1500 cm$^{-1}$ are respectively assigned to asymmetric and symmetric stretching in OCO and vibration of C=C in benzene ring[11,18-19,23]. This further confirms that the amino functional group or small amount of GO does not critically disturb the formation of UiO-66 and the spectra are dominated by the skeletal modes of the UiO-66[15]. Besides the main peaks, because the presence of post-synthesized functional group, features of the essential spectrum could be observed. UiO-66-NH$_2$ and UiO-66-NH$_2$/GO demonstrate three additional peaks at 1255, 1370 and 1622 cm$^{-1}$. The first two peaks are assigned to C–N stretching in the aromatic carbon and nitrogen, while the last peak belongs to bending vibration of N–H[11]. Moreover, compared to UiO-66-NH$_2$/GO, the amino group in UiO-66-NH$_2$ has apparent character in high frequency spectrums. It suggests the primary aromatic amino groups demonstrated two medium absorbances corresponding to the asymmetric and symmetric stretching. One is around at 3507

cm$^{-1}$ and the other is about at 3384 cm$^{-1}$ [26] and their average is about 3430.81 cm$^{-1}$. The changes in the peak position and shape can indirectly prove that the original UiO-66 has been impacted by the introduction of amino functional group and GO[18].

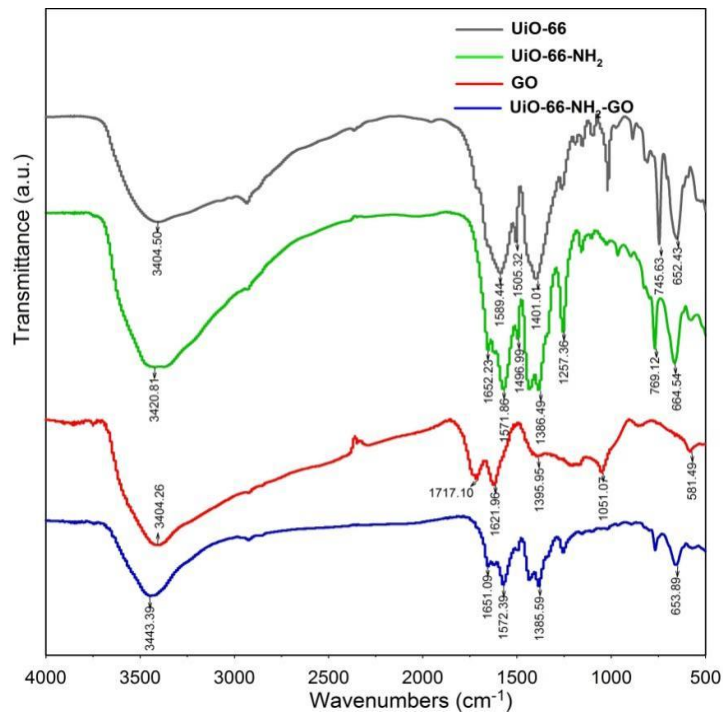

Fig. 9  FTIR Spectrum of four materials

### 3.1.4 Nitrogen adsorption-desorption isotherms analysis

To investigate the effects of amino group and GO modification on surface area and pore structure in UiO-66 framework, the adsorption-desorption isotherms with nitrogen gas at 77K (Fig.10) was conducted and the pore distribution of each synthesized product was obtained as shown in Fig.11. The sorption isotherms of UiO-66, UiO-66-NH$_2$ and UiO-66-NH$_2$/GO (Fig.10) were catalogued into IV Type. The H$_2$ hysteresis loop in IUPAC classification indicates a harmonic multilayer adsorption on uniformed surface of the solid. All the N$_2$ sorption isotherms (specific pressure P/P$_0$ =0~0.1) reveal that the adsorption of N$_2$ is increasing rapidly. It displays a strong interaction for all materials toward N$_2$ by micropore filling. Comparing to UiO-66 (Figure 11), the isotherm of UiO-66-NH$_2$ has shown a Type IV isotherm at pressure P/P$_0$ around 0.64. After the introduction of GO, a more apparent adsorption hysteresis ring occurs when specific pressure P/P$_0$ is around 0.44. This indicates that a certain amount of amino functional group and GO could participate in the coordination of the metal center, generate defect sites on the crystal framework decomposing under high temperature and lead to the creation of mesopores in the crystal[26].

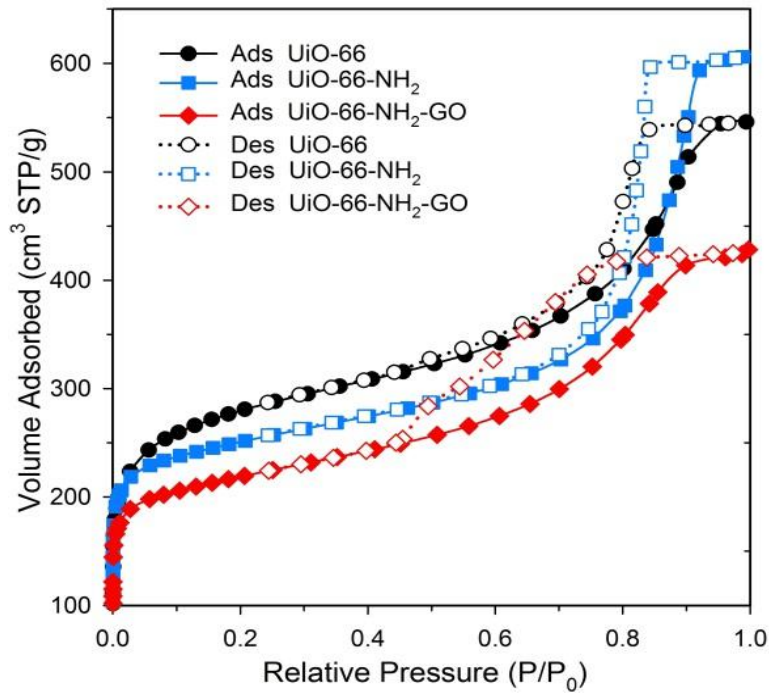

Fig. 10 N₂ adsorption Isotherms of adsorbents at 77 K

According to Table 4, (1) the incorporation of amino group on UiO-66 decreased the surface area from 990.06 m²/g to 925.21 m²/g and further reduction to 801.90m²/g results from the introduction of GO. The Amino group and GO blocked some pores on original UiO-66, which made the diffusion rate of liquid nitrogen in hybrid network lower, thus reducing the specific surface area. (2) From the pore distribution data of materials, the micropore volume for UiO-66-NH₂ has subtly reduced but the mesopore volume has significantly enlarged compared to UiO-66. However, the mesopore and micropore volume for UiO-66-NH₂/GO have both decreased dramatically in comparison to UiO-66 and UiO-66-NH₂. This indicates that the porosity of UiO-66 affected positively through the incorporation of amino group. Thus improve the accessibility toward the active sites of UiO-66[19,24]. Moreover, because 2-4nm pores benefit the diffusion and adsorption of PFOA, UiO-66-NH₂ gain an advantage in pore structure for the adsorption of PFOA. Nevertheless, the introduction of GO would greatly minimize the volume of mesopores and hinder adsorption. (3) In respect to total pore volume, UiO-66-NH₂ has the maximum(0.9377cm³/g) and UiO-66-NH₂/GO the minimum (0.6577 cm³/g). The result demonstrates the incorporation of NH2 group could increase the pore size and volume of UiO-66 and provides a higher adsorption capacity, corresponding to previous studies[11]. Compared with the original UiO-66 without any modification, the incorporation of GO damages the structure of UiO-66 by attaching GO molecules to the pores of UiO-66-NH2, which decrease the useable pores and increase the total weight of MOF and lead to a significant decrease in pore width, pore volume and specific surface area[18,23,27].

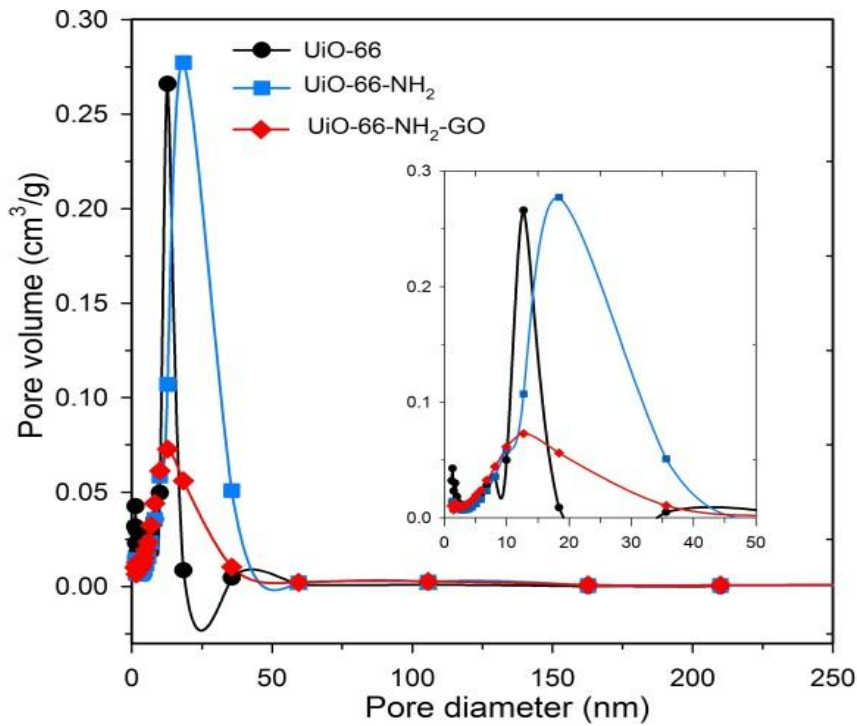

Fig. 11  Pore distribution of three adsorbents

Table 4 BET results of three adsorbents

| Material Name | Surface Area ($m^2 \cdot g^{-1}$) | pore volume ($cm^3 \cdot g^{-1}$) | | | |
|---|---|---|---|---|---|
| | | Pore width <2nm | Pore width 2~50nm | Pore width >50nm | Total |
| UiO-66 | 990.06 | 0.22 | 0.51 | 0.00 | 0.73 |
| UiO-66-NH$_2$ | 925.21 | 0.11 | 0.63 | 0.01 | 0.75 |
| UiO-66-NH$_2$/GO | 801.9 | 0.09 | 0.40 | 0.01 | 0.50 |

### 3.1.5 Contact angle analysis

The water contact angle of UiO-66, UiO-66-NH$_2$ and UiO-66-NH$_2$/GO and GO is 10.5°, 146.7°, 40.1° and 28.2°, respectively. Accordingly, the introduced amino group could significantly increase the hydrophobicity of the original UiO-66. However, the hydrophobic effect was greatly weakened after the further introduction of GO, due to its oxygen-rich component.

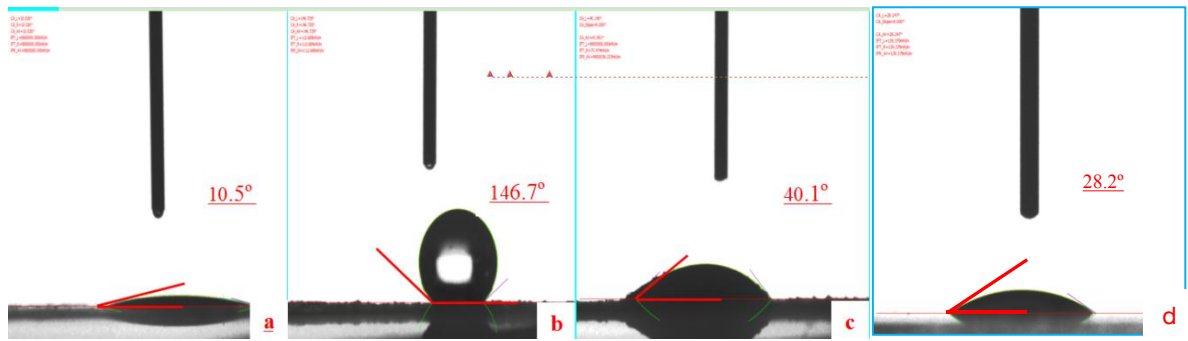

**Fig. 12  Contact angle of synthesized products (a.UiO-66; b.UiO-66-NH$_2$; c. UiO-66-NH$_2$/GO; d. GO)**

### 3.1.6 Zeta Potential analysis

The table 5 suggests that all materials carry positive charge, in which UiO-66 reaches the highest with 22.4mV and UiO-66-NH$_2$/GO reaches the lowest with 12.9mV. The data indicates that the modification of amino groups and GO reduced the electrostatic interactions towards anions due to the basicity caused by the ionization of the amino group in water. All of synthesized materials demonstrated similar trend of zeta potential, where the state of positive charge appears at a low pH and changes to the state of negative charge when the pH is increased[11], which accounts for the zero charge point (pH$_{PZC}$) of all materials is higher than 7. In this research, the adsorption experiments were conducted at pH=3±0.1, which suggested that all three materials were in the same state of charge.

**Table 5  Zeta Potential of three adsorbents**

| material name | UiO-66 | UiO-66-NH$_2$ | UiO-66-NH$_2$/GO |
| --- | --- | --- | --- |
| Zeta potential（mV） | 22.4 | 17.6 | 12.9 |

## 3.2 Adsorption studies

### 3.2.1 Adsorption efficiency and capacity

The adsorption performances of PFOA onto three sorbents are shown in Table 6. The UiO-66-NH$_2$ demonstrated higher the adsorption efficiency of PFOA. The removal rate reached 80% after 3 hours reaction and the adsorption capacity was 952.5 mg/g. Comparably, the adsorption efficiency of UiO-66-NH$_2$/GO for PFOA is lower, achieving 70% in 3 hours with an adsorption capacity of 800.0 mg/g. It suggests that the introduction of GO hindered the adsorption of PFOA on the UiO-66-NH₂. From the results, it could be safely stated that UiO-66-NH₂ is the best adsorbent among the three.

Table 6  Adsorption Efficiency and Capacity of three adsorbents

| Material Name | UiO-66 | UiO-66-NH₂ | UiO-66-NH₂/GO |
|---|---|---|---|
| Adsorption Efficiency (%) | 62 | 80 | 70 |
| Adsorption Rate (mg/g) | 648.5 | 952.5 | 800 |

### 3.2.2 Adsorption kinetics

Two kinetics models (Pseudo first-order kinetic and pseudo second-order kinetic model) were applied to analyze PFOA adsorption of UiO-66-NH₂. The pseudo-second order kinetics model were fitted the date better than that of the pseudo-first-order kinetics (Fig.13). The calculated qe value of second order model was closer to the actual value. The result verifies that the UiO-66-NH₂ adsorption of PFOA follows the pseudo-second-order kinetics. Therefore, the adsorption of PFOA onto UiO-66-NH₂ was mainly controlled by chemisorption.

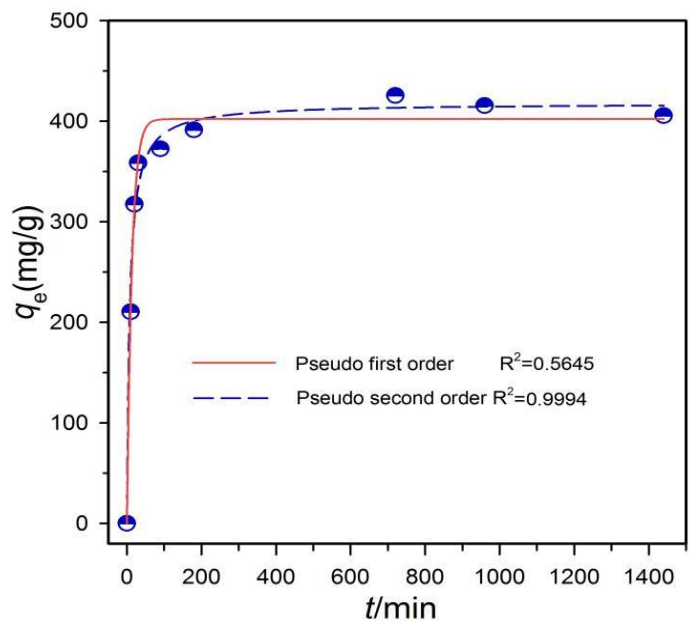

Fig. 13  Adsorption kinetics of UiO-66-NH₂

### 3.2.3 Adsorption isotherms

To detect the adsorption behavior of UiO-66-NH₂ and maximum adsorption capacity, 0.01g of the adsorbent was added to 5~300 mg•L⁻¹ PFOA solution (25°C, pH=3±0.1) for 24 hour to ensure the reach of equilibrium (Fig.14). Compared with the Langmuir model ($R^2$= 0.969), the Freundlich model ($R^2$= 0.986) has a higher fitting. Therefore the adsorption of PFOA onto UiO-66-$NH_2$ accords with the theory of multi-layer adsorption. In other words, the adsorbent layer has multilayers with different physical and chemical incentives, and correspondent chemisorption and physisorption present on a surface that have unequal active sites presents[18]. Based on the Freundlich equation, n>1 means the adsorption is favorable, and a large n value suggests a more plausible adsorption, matching with a stronger adsorption drive force. The value of n for UiO-66-$NH_2$ is 2.11, indicating a meticulous absorbability[18]. Furthermore, a stronger adsorption capacity of UiO-66-$NH_2$ for PFOA was 1653 mg/g (maximal adsorption capacity ) according to Freundlich model, apparently higher than the adsorption capacity （1269mg/g）of 3D hierarchically microporous biochar [12].

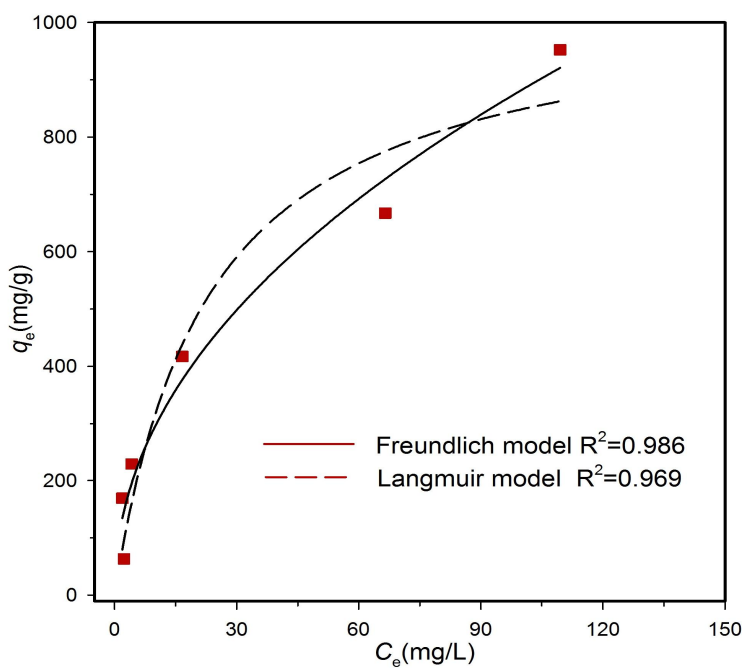

Fig. 14 Adsorption Isotherms of UiO-66-$NH_2$

### 3.3 Adsorption mechanisms

According to the adsorption kinetics and isotherms, the PFOA adsorption process on UiO-66, UiO-66-$NH_2$ and UiO-66-$NH_2$/GO is complicated and multi-factor controlled. To begin with, compared to UiO-66, modified material UiO-66-$NH_2$ has a reduced surface area, but with a higher pore volume and mesoporous volume, which is more conducive for the dispersion of 1.2nm PFOA molecule[28]from the solution to the pore exterior surface, pore space and pore interior surface of UiO-66-NH$_2$, contributing to its high absorbability. In comparison, UiO-66-NH$_2$/GO has a markable reduction in surface area, pore volume, mesopore and micropore size from UiO-66, which structurally hinder the adsorption of PFOA. In addition, the adsorption experiment is performed under acidic situation (pH=3.0±0.1) where PFOA carries negative charges, and hydronium ion(H$_3$O$^+$) covers the adsorbent surface. This could result in the acid and base interaction between the hydronium ion on each adsorbent surface and the lone pair electrons on PFOA. Furthermore, three adsorbents carry positive charge under acidic situation (pH=3.0±0.1) and the strength of the electronegativity of UiO-66-NH$_2$ is between UiO-66 and UiO-66-NH$_2$/GO, causing a stronger electrostatic interaction towards the negative charged PFOA molecule compared with UiO-66-NH$_2$/GO but weaker than UiO-66.

Beyond that, the presence of different functional groups would affect the number of hydrogen bonds that the material could establish with PFOA, lead to an effect on the adsorption capacity. With respect to UiO-66 surface, its inherent organic ligands could connect more hydrogen bonds with PFOA, rendering better absorbability. In terms of UiO-66-NH$_2$, the amino group is the nucleophile group, enhancing the electron density of organic ligand, the amino group facilitates the generation of hydrogen bond, strengthens its effect than on UiO-66. The GO contains oxygenic function groups such as carboxyl group, epoxy group and carboxylic acid group, et al. promote the adsorption of PFOA on UiO-66-NH$_2$/GO by fostering intensified hydrogen bond. At last, UiO-66-NH$_2$ possesses the strongest hydrophobic properties among the three adsorbents. The introduction of GO functional group not only diminished the pore adsorption interaction but also weakened its hydrophobicity.

In general, we propose possible adsorption mechanisms that pore filling interaction, acid and base interaction, hydrogen bonding, electrostatic interaction and hydrophobic interaction are played the main role in PFOA adsorption onto MOFs [14]. The electrostatic interaction and pore adsorption interaction mainly contributes to the adsorption mechanism of UiO-66; on the other hand, after the modification of UiO-66-NH$_2$ by GO functional group, decreasing its hydrophobicity and pore adsorption interaction, the composite material relies heavily on hydrogen bonding for adsorption, leads to a smaller adsorption capacity from original UiO-66-NH$_2$. As a result, after the introduction of amino group, the increase in the mesopore size and improvement in hydrophobicity are the two main attributes for the stronger absorbability of PFOA than the original UiO-66. The adsorption isotherm analysis revealed that physisorption and chemisorption of PFOA on UiO-66-NH$_2$ are involved in. The adsorption kinetics indicated a controlled chemisorption mechanism for adsorption process. Therefore, it could be proposed that the hydrophobic property can be the main driving force for PFOA adsorption of UiO-66-NH$_2$ (Fig. 15).

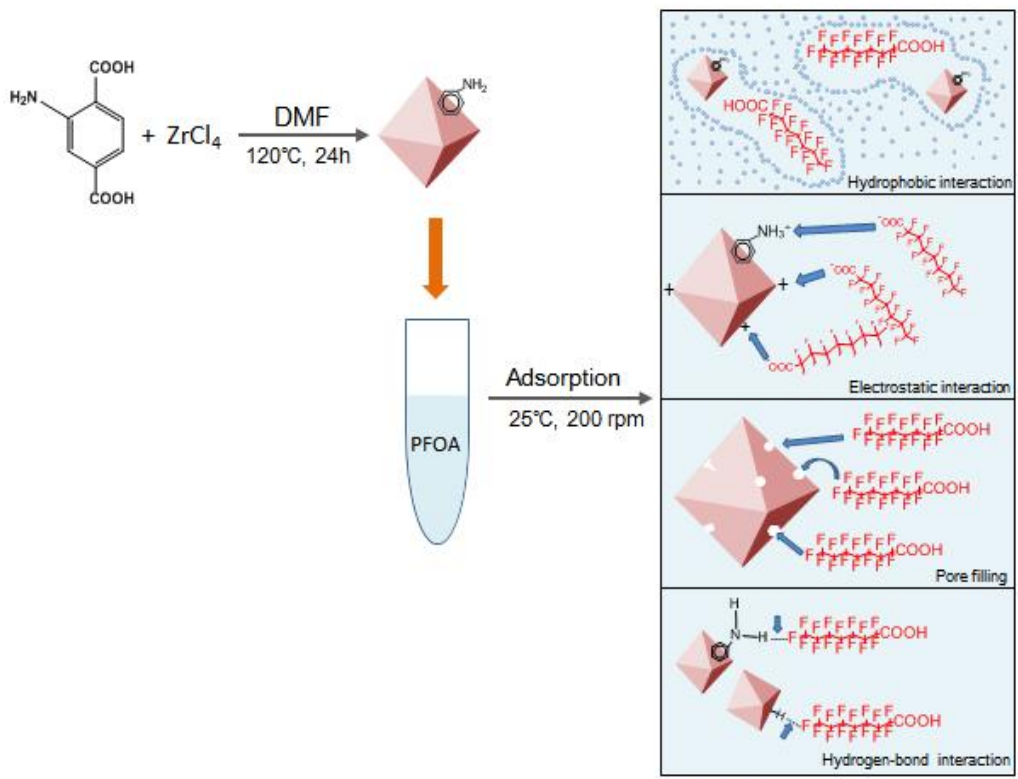

Fig. 15 Adsorption mechanisms of UiO-66-NH$_2$

**4 Conclusion**

  This research mainly focused on the hydrophobic effect of PFOA adsorption by firstly introducing the amino group to UiO-66 and further introducing GO to modify composite material. Based on the successful synthesis of Zr-based MOFs through a simple hydrothermal method, the properties and mechanism of PFOA adsorption on each material are explored. The result indicates that amino group enlarges the pore size as well as exhibiting intensive hydrophobicity, escalating the adsorption capacity of PFOA (maximum capacity of 1653 mg/g). Pseudo-second-order and Freundlich model well fitted the adsorption process and behavior of PFOA on UiO-66-NH$_2$. This work provides the method of introducing amino group to Zr-based MOFs multifunctional hybridized material to strengthen the hydrophobic properties in purpose of removing PFOA efficiently. More importantly, this work provides verification for a promising wastewater management approach in which the designability of MOFs(such as Zr-MOFs) was utilized to introduce particular functional group for modifying MOFs according to the properties of the pollutant, that is to tailor specialized adsorbent for enhancing the adsorption elimination of contaminants.

**5 Suggestion for further research**

The results of this research proved that the modification of amino group on UiO-66 could magnificently strengthen hydrophobicity and therefore increase the adsorption capacity of PFOA. The suggestion for the further research is to introduce more amino functional groups to UiO-66 for further improving its hydrophobic effects, for example, to decorate the post-synthesized UiO-66 MOFs with urea functional group which has two amino groups[29], and investigate the adsorption capacity for PFOA, so that provide technological support for improvement in the elimination of PFOA.

# Acknowledgement

I sincerely give thanks to the financial support offered by the *Independence Science Research Project* in George School, PA and *Chongqing Adolescents Innovative Talent Project*.

I am also grateful for the support Prof. Zeng from College of Environment and Ecology, Chongqing University and her research team. I managed to read through articles, choose topics, plan, operate experiments, processed data and write report on my own with Prof. Zeng's patient guidance and valuable advice.

I whole heartfully thank for the excellent experimental platform provided by Key Laboratory of Three Gorges Reservoir Area, Ministry of Education, Chongqing University, which promotes the success of the research work.

Besides, Brian, the instructor in the *Independence Science Research Project* in George School is so supportive and helpful when revising me essay. His constructive advice and help certainly facilitate my essay writing.

Finally, I appreciate the analyzing and testing center in Chongqing University and Lie Pu Technology in Suzhou City, Jiangsu Province for their hospitality and timely service. With their assistance, I could finish PXRD, SEM, BET, FT-IR, Contact Angle, Zeta Potential and PFOA concentration test.